\documentstyle[fleqn,twoside,gc]{article}

\newcommand{\be}{\begin{equation}}
\newcommand{\ee}{\end{equation}}
\newcommand{\bea}{\begin{eqnarray}}
\newcommand{\eea}{\end{eqnarray}}

\heads{Akiko NOGUCHI and Akio SUGAMOTO}
      {Dynamical Origin of Duality between Gauge Theory and Gravity}

\begin{document}
\twocolumn[

\Title{Dynamical Origin of Duality between Gauge Theory and Gravity}


   \Author{Akiko Noguchi and Akio SUGAMOTO}   
          {Department of Physics and Graduate School of Humanities and Sciences, \\
Ochanomizu University, 
Tokyo 112-8610, Japan}              

\Abstract
    {Dynamical origin of duality between gauge theory and gravity is studied using the dual transformation and the formation of graviton as a collective excitation of dual gauge bosons.
In this manner, electric-magnetic duality in gauge theory is reduced to the duality between gauge theory and gravity.
}



]  

\section{Gravity as collective excitation of dual gauge fields}
 
We are very happy to submit this paper to the special issue of Gravitation and Cosmology on the occasion of 70 years anniversary of Faculty of Physics and Mathematics of Tomsk State Pedagogical Univsesity.

It is popular now that by the AdS/CFT correspondence developed by J. Maldacena~\cite{Maldacena}, gauge theory and gravity become dual with each other.  In other words, the strong coupling regime of the former theory corresponds to the weak coupling regime of the latter, and  {\it vice versa}.  In this corresponcence D-branes and extra dimension play essential roles.  This AdS/CFT correspondece may be considered within the context of the 't Hooft-Mandelstam duality~\cite{'t Hooft-Mandelstam}.   

In the late 70's  and the early 80's, similar correspondence was known in which the closed string theory (with the Kalb-Ramond field as a gauge field of string) is dual to the gauge theory (being massive with or without Higgs field), and the extension of membranes is also considered~\cite{Nambu, AbelianD-T, non-AbelianD-T, Membrane}.  

Malcacena's  AdS/CFT correspondence is, of cource, the more sophisticated, and provides an extremely powerful tool in studying the realistic hadron physics in QCD.  Examples of such study can be found in a review article~\cite{Ooguri} and in a recent study of pentaquark baryons using the AdS/CFT correspondence~\cite{BKST}. 

In this paper, we study an origin of the dualiy between gauge theory and gravity (AdS/CFT) within the local field theory using the old-fashioned duality.

The idea which our study is based on is as follows.  Starting from a gauge theory and applying a dual transformation to it, we obtain a dual theory.  (We may start from a manifestly self-dual theory also.)  If the gauge coupling of the original theory is $e$, then the gauge coupling of the dual theory is $g=2\pi/e$.  When $e$ is small, $g$ is large, so that bound states (or collective excitations) are formed in the dual theory by the exchange of strongly coupled dual gauge bosons.  Among them we have a graviton as the collective excitaion of dual gauge bosons.  Then, the graivity theory becomes dual to the original gauge theory, since the former is the low energy manifestation of the dual gauge theory in the strong coupling regime.  

This idea is quite consistent with the fact that in string theories, a gauge boson is represented by an open string's  mode, while a graviton is by a closed string's  mode, so that a closed string can be considered as a bound state of two open strings.

We examine this idea in the usual $U(1)$ gauge theory, and also in its manifestly self-dual formulation by Zwanziger~\cite{Zwanziger}. 
More detailed description of this paper can be found in the master theis written by one of the authors (A.N.)~\cite{Noguchi}


\section{Duality between U(1) gauge theory and gravity}
We start from the $U(1)$ gauge theory with a coupling $e$.  The parition function of this theory reads,
 \begin{eqnarray}
 Z[J] &\propto&
       \int {\cal D} A_{\mu} \exp \left\{ i \int d^4 x
       \left[ - \frac{e^2}{4} F_{\mu \nu} F^{\mu \nu} \right. \right. \nonumber \\
&&\left. \left. + \frac{1}{2} F_{\mu\nu} J^{\mu\nu} \right] \right\}, \label{ua}
   \end{eqnarray}
where $F_{\mu\nu}=\partial_\mu A_\nu - \partial_\nu A_\mu$ is the field strength of the gauge field $A_\mu$, and $J^\mu$ is an external source.

Using an auxilliary field $W_{\mu \nu}$, the partition function becomes~\cite{Nambu, AbelianD-T},
   \begin{eqnarray}
    Z[J]&\propto&\int {\cal D} A_{\mu} \int {\cal D} W_{\mu \nu} \nonumber \\
   && \times \exp \left\{i \int d^4 x \left[-\frac{e^2}{4} W_{\mu \nu} W^{\mu \nu}+\frac{1}{2} \tilde{W}_{\mu \nu} F^{\mu \nu} \right. \right. \nonumber \\
&& \left. \left. +\frac{1}{2} F_{\mu\nu} J^{\mu\nu} \right] \right\}, \label{uaw}
   \end{eqnarray}
where
   \begin{eqnarray}
     \tilde{W}_{\mu\nu} &=& \frac{1}{2} \epsilon_{\mu\nu\lambda\rho} W^{\lambda\rho},~~~ 
     (\epsilon_{0123}=-\epsilon^{0123} = 1).
   \end{eqnarray}
Path integration over $A_{\mu}$ leads  (\ref{ua}) to 
\begin{eqnarray}
&&\int {\cal D} B_{\mu} \exp \left\{ i \int d^4 \left[-\frac{1}{4g^2} G_{\mu \nu} G^{\mu \nu} + \frac{1}{2} \tilde{G}_{\mu \nu} J^{\mu \nu} \right. \right. \nonumber \\
&& \left. \left. +\frac{1}{2} J_{\mu \nu} J^{\mu \nu} \right] \right\},      \label{uab}
\end{eqnarray}
 where $G_{\mu \nu} \equiv \partial_\mu B_\nu - \partial_\nu B_\mu$ is the field strenght of the dual gauge field $B_{\mu}$, and $e g=2\pi$ holds.

The method to obtain the action of a bound state as a collective excitaion was known in the Nambu-Jona-Lasinio model, and in the gauge and gravity models~\cite{NJL, CollectiveExcitation}.  

In order to keep the general covariance, the method proposed by Akama in~\cite{Terazawa}, using the dimensional regularization, is very useful.   
To introduce the graviton into the free Lagrangian of the dual gauge fields in (\ref{uab}) as a collective excitation of gauge bosons, the original Lagrangian,
\begin{eqnarray}
     {\cal L}^{(0)} = - \frac{1}{4g^2} G_{\mu\nu}G^{\mu\nu}, 
\end{eqnarray}
is to be modified general covariantly, by introducing a background metric $g_{\mu\nu}(x)$ as the collective coordinate of graviton.  Then, we obtain
\begin{eqnarray}
     {\cal L}^{(1)} = -\frac{1}{4g^2} \sqrt{g}g^{\mu\lambda}g^{\nu\rho} G_{\mu\nu}G_{\lambda\rho}. 
\end{eqnarray}
If the meric is expanded around the flat space,
\begin{eqnarray}
g_{\mu\nu}=\eta_{\mu\nu}+h_{\mu\nu}, 
\end{eqnarray} 
the fluctuation $h_{\mu\nu}$ describes a graviton excitation.

Next task is to eliminate the original dual gauge fields by path integration.  Then, we obtain the effective action of gravity which is the low energy effective theory of the dual gauge theory.  This is not difficult to perform.  Follwoing Akama, we arrive at
 \begin{eqnarray}
    {\cal L}_{\rm{eff}} &=& \sqrt{-g} \left[-\frac{1}{2} \Lambda^4-\frac{1}{8\pi}\Lambda^2 R \right. \nonumber \\
&& \left. -\frac{1}{480\pi^2} \ln \Lambda^2 (R^2 - 3 R_{\mu\nu}R^{\mu\nu}) \right],
  \end{eqnarray}
where $\Lambda$ is the ultraviolet cutoff.  Here, the Newton constant is negative, giving the repulsive gravitational interaction, but this can be remedied by introducing a number of dual fermions and scalars into the theory.  

The important relation given by Terazawa {\it et al.} in~\cite{Terazawa} between the Newton constant and the fine structure constant, is to be modified, since in the present context, the graviton is the collective excitaion not of the usual gauge bosons, scalar and fermions, but of the "dual"gauge bosons, scalars and fermions.  Therefore, the fine structure constant $\alpha$, and the number of fields are to be replaced by the dual couping constant $\tilde{\alpha}$, and the number of the dual fields, respectively.  Here, $\alpha$ and $\tilde{\alpha}$ are inversely related, following $\alpha\tilde{\alpha}=1/4$.

Green functions of the original gauge theory, those of the dual gauge theory and those of the dual gravity theory are related as follows:
\begin{eqnarray}
     &&\langle F_{\mu\nu}(x) F_{\lambda\rho}(y) \cdots\rangle_{\rm original ~gauge} \nonumber \\
    &&=\frac{1}{g^2}\langle\tilde{G}_{\mu\nu}(x) \tilde{G}_{\lambda\rho}(y) \cdots\rangle_{\rm dual~gauge} \\
&&+\frac{1}{2g^2}\delta^{(4)}(x-y)(\eta_{\mu\lambda}\eta_{\nu\rho}-\eta_{\mu\rho}\eta_{\nu\lambda}\langle \cdots \rangle_{\rm dual ~gauge}. \nonumber \\
&&\langle \frac{1}{g^2}G_{\mu\nu}(x)G_{\lambda\rho}(y) \cdots\rangle_{\rm dual~gauge} \\  
&&=\langle -P_{\mu\nu}^{\alpha}(x)O^{-1}_{\alpha\beta}(x, y)P_{\lambda\rho}^{\beta}(y) \cdots \rangle_{\rm dual~gravity}, \nonumber
\end{eqnarray}
where the operators in the dual gravity theory are given by
\begin{eqnarray}
P_{\mu\nu}^{\alpha}&=&\partial_{\beta}\sqrt{-g}P^{\alpha\beta}_{\mu\nu}, \\
O_{\alpha\beta}&=&\partial_{\gamma}\sqrt{-g}P^{\alpha\gamma}_{\beta\delta}\partial_{\delta},  \\
P_{\mu\nu, \alpha\beta}&=&g_{\mu[\alpha}g_{\nu\beta]}-g_{\nu[\alpha}g_{\mu\beta]}.
\end{eqnarray}
These relations give the dual correspondence between the gauge theory and the gravity in terms of the corresponding Green functions.


  \section{Duality between Zwanziger's U(1) gauge theory and closed string theory}
The Zwanziger's model is the manifestly self-dual formulation of $U(1)$ gauge theory~\cite{Zwanziger}.

If we introduce the collective coordinate of exciting graviton into the Zwanziger's action, we obtain   \begin{eqnarray}
&& S^{\rm{ZW}}=\int d^4 x \left(-\frac{1}{2} n^{\mu} n^{\lambda}
               \sqrt{-g} g^{\nu\rho}\right) \nonumber \\
&&\times\left(F_{\mu\nu} F_{\lambda\rho} + G_{\mu\nu} G_{\lambda\rho}+ iF_{\mu\nu} \tilde{G}_{\lambda\rho} - iG_{\mu\nu}\tilde{F}_{\lambda\rho}\right).
   \end{eqnarray}
Here, a constant unit vector $n^{\mu}$ denotes the direction of Dirac strings which are assumed to be frozen in time, being  displayed in parallell, and 
 \begin{eqnarray}
            F_{\mu\nu}&=&\partial_{\mu}A_{\nu}-\partial_{\nu}A_{\mu}, \\
           G_{\mu\nu}&=&\partial_{\mu}B_{\nu}-\partial_{\nu}B_{\mu}, \\
           \tilde{F}_{\mu\nu}&=&\frac{1}{2} \varepsilon_{\mu\nu\lambda\rho}
                               F^{\lambda\rho}. 
\end{eqnarray}

What we have to do next is to eliminate the gauge fields $A_{\mu}$ and $B_{\mu}$ by path integration.  
Before doing this, it is better to modify the frozen direction $n^{\mu}$ of the Dirac string to be the dynamical coordinate of string.  For this purpose, we perform the following replacement;
\begin{eqnarray}
     n^{\mu}n^{\nu} \to \int d\tau d\sigma 
      \left[\left(\frac{\partial X^\mu}{\partial \tau}\right)^2
      -\left(\frac{\partial X^\nu}{\partial \sigma}\right)^2\right],
   \end{eqnarray}
where $X^{\mu}(\tau, \sigma)$ represents the world sheet of the Dirac string, parametrized by two parameters, $\tau$ and $\sigma$.  To eliminate the gauge fields $A_{\mu}$ and $B_{\mu}$, the axial gauge fixing and its Feynman rules are useful~\cite{Lape}.

Then, we obtain the lowest non-trivial action for the dynamical Dirac string;
   \begin{eqnarray}
     S_{\rm{eff}} &=& C_0 \Lambda^4 \int d^4 x \int d\tau d\sigma
                 \sqrt{-g}g_{\mu\nu} (X(\tau,\sigma)) \nonumber \\
             & &\times\left[\left(\frac{\partial X^\mu}{\partial \tau}\right)^2
                 -\left(\frac{\partial X^\mu}{\partial \sigma}\right)^2\right].
   \end{eqnarray}
This action describes the dynamics of Dirac string moving in the curved background space, and gives, at the same time, the dual theory of the $U(1)$ gauge theory.
 \section{Non-Abelian theory}
To carry out the dual transformation in non-Abelian gauge theories, one method is to follow~\cite{Nambu, non-AbelianD-T}.  So far this was a rather complicated method, but very recently a new mathematical concept is introduced into this method~\cite{NAKalbRamond}.  This new direction is expected to give the clearer understanding of the non-Abelian dual transformation.

The other method is to use the non-local loop variables from the beginning~\cite{Chan, Lape}.  In this context, the non-Abelian Zwanziger type action is also known~\cite{Lape}
 \begin{eqnarray}
    && S^{NAZW}= \nonumber \\
    &&-\frac{1}{4\pi} \int {\cal D}\xi^{\mu}(s) ds 
        \{(\dot{\xi}^\mu(s)\bf{F}_{\mu\nu}(x))
        (\dot{\xi}^\lambda(s)\bf{F}^\nu_{\lambda}(x))\dot{\xi}^{-2}
        \nonumber \\
       &&+(\dot{\xi}^\mu(s)\bf{G}_{\mu\nu}(x))
        (\dot{\xi}^\lambda(s)\bf{G}^\nu_{\lambda}(x))\dot{\xi}^{-2}
        \nonumber \\
       &&+(\dot{\xi}^\mu(s)\bf{F}_{\mu\nu}(x))
        (\dot{\xi}^\lambda(s) \tilde{\bf{G}}^\nu_{\lambda}(x))\dot{\xi}^{-2}
        \nonumber \\
       &&+(\dot{\xi}^\mu(s)\bf{G}_{\mu\nu}(x))
        (\dot{\xi}^\lambda(s) \tilde{\bf{F}}^\nu_{\lambda}(x))\dot{\xi}^{-2}\},
   \end{eqnarray}
where $\xi^{\mu}(s)$ represents the direction of string and sum over the string's shape $\int {\cal D}\xi^{\mu}(s)$ is introduced.  The tilde here represents the "dual", but is not the simple "Hodge dual" in this non-Abelian case.

It is an intereting problem to derive the dual gravity theory, starting  from this non-Abelinan Zwanziger type action.

[Note added]
In the non-Abelian duality, we have to refer to the pioneering works by M. B. Halpern, {\it Phys.  Rev.} {\bf D16}, 1798 (1977); {\it i.b.d.} {\bf D16}, 3515 (1977); {\it i.b.d.} {\bf D19}, 517 (1979), in addition to \cite{non-AbelianD-T}.

\Acknow
{One of the authors (A.S.) gives his sincere thanks to Professor Y. Nambu for valuable discussions on the basic idea of this paper in the summer of 2002 in Chicago.  He is also grateful to Rector Obukhov and Sergei Odintsov of Tomsk State Pedagogical University for promoting international excahnge program and collaborations.}

\end{document}